# Capillary-Assisted Printing of Droplets at a Solid-Like Liquid-Liquid Interface

Anshu Thapa[1], Robert Malinowski[1], Matthew O. Blunt[1], Giorgio Volpe[1*], Joe Forth[1,2,3*]

[1]Department of Chemistry, University College London, 20 Gordon Street, London, WC1H 0AJ, UK
[2]Department of Chemistry, University of Liverpool, Crown Street, Liverpool, L69 7ZD, UK
[3]Department of Physics, University of Liverpool, Oxford Street, Liverpool, L69 7ZE, UK

*Email: j.forth@liverpool.ac.uk; g.volpe@ucl.ac.uk

**Capillary forces guide the motion of biomolecular condensates, water-borne insects, and breakfast cereal. These surface-mediated interactions can be harnessed to build units into materials with exotic properties deriving from mesoscale structure. Droplets are promising building blocks for these materials, finding applications in tissue engineering, adaptive optics, and structural colour. However, the instability of water droplets at many liquid-liquid interfaces hampers the use of capillarity for the assembly of droplet-based materials. Here, we use nanoparticle surfactants to form solid-like oil-water interfaces at which aqueous droplets sit for extended periods. We find that microlitre-sized droplets at these interfaces attract each other over millimetric scales. We rationalize this interaction with a modified theory of capillarity. Applying printing methods allows us to finely control initial droplet positions, from which they self-assemble into cellular materials. Finally, by functionalising the interface with gold nanoparticles, we use plasmon-assisted optofluidics to manipulate these droplet-based materials with temperature gradients.**

Materials made of liquids offer enormous potential for building shape-shifting devices that can reconfigure themselves on-demand.[1] Such systems are inherently self-healing, meaning they can be endlessly reshaped and reprocessed into a myriad of complex structures.[2] The liquids within these structures conduct solutes, allowing for directed transport of materials.[3] The difference in physicochemical properties between two immiscible liquids results in partitioning and compartmentalisation of reagents and reactants. Communication between compartments can lead to the emergence of complex collective droplet behaviours such as flocking, chasing, and actuation.[4] These ideas have been explored to make microfluidic chips that can be reconfigured *in operando* and active droplets that drive themselves from equilibrium by consumption of chemical fuel.[5,6] Adding advanced chemistries into liquid materials can in turn lead to applications in reconfigurable optics,[7] dynamic materials patterning,[8] and soft robotics.[9]

The interface between two liquids further enables the surface tension-driven assembly of functional materials.[10] At the simplest level, materials adsorbed at the liquid-liquid interface enhance the stability of droplets against coarsening and coalescence.[11] Solid-like adsorbed layers of particles and polymers can further be used to shape liquids into complex structures ranging from fibrils to tortuous, bicontinuous systems.[12] More recently, liquid-liquid interfaces have been harnessed to incorporate biological materials and nanoparticles, enabling complex thermal response,[13] magnetic manipulation,[14] or size- and chemical-selective transport between liquid compartments.[15]

The simple design rules above can be connected to self-assembly and additive manufacturing techniques to produce time-evolving systems made from droplets and liquids.[16] DNA-functionalisation of the liquid-liquid interface allows for molecular recognition and, hence, programmed self-assembly of specific droplet architectures.[17] Gradients in surface tension lead to Marangoni stresses, which can be used as an external handle to manipulate droplet assembly.[18,19] Droplet printing enables the fabrication of tissue-like materials that undergo structural transformation due to mechanical stresses that are spatially patterned into the system.[20] Capillary forces have been harnessed in this context to direct the assembly of anisotropic solid particles and template the formation of optical metamaterials.[21,22] To a limited degree, these surface-mediated interactions, which derive from the inherent stress at the interface between two immiscible phases, can also be used to assemble droplet-based systems.[23] A number of approaches have been developed to place water droplets at, or hang them from, the air-water interface;[24,25] however, these methods allow only a limited palette of materials and chemistries to be incorporated into the system. What is desirable is a system that allows aqueous droplets to be placed at the oil-water interface for long times, allowing capillary forces to be harnessed to produce a wholly liquid, cellular material.

In this work, we assemble a solid-like layer of nanoparticle surfactants at the oil-water interface at which aqueous droplets can be placed for extended periods of time. Capillary interactions between the droplets lead to inter-droplet attractions at millimetric separations. We quantify these interactions, capturing the scaling of the size, separation, and composition of the droplets with their speed by modifying the classical theory of capillary forces to account for the drag between the droplets and the solid-like oil-water interface. We combine this capillary-force-driven assembly with printing methods to spatially define the initial locations and composition of the droplets, yielding a technique we term 'capillary-assisted printing', in which the droplets then build themselves into a range of structures. Finally, we leverage the flexibility in composition of the nanoparticle surfactant assembly to functionalise the solid-like oil-water interface with gold nanoparticles, giving us a thermoplasmonic handle with which to manipulate the droplets and their assemblies at the interface. Upon illuminating the interface using a laser with a wavelength at the plasmonic resonance of the nanoparticles, strongly localised heating drives motion of the droplets, allowing us to manipulate multi-droplet structures along controlled trajectories.

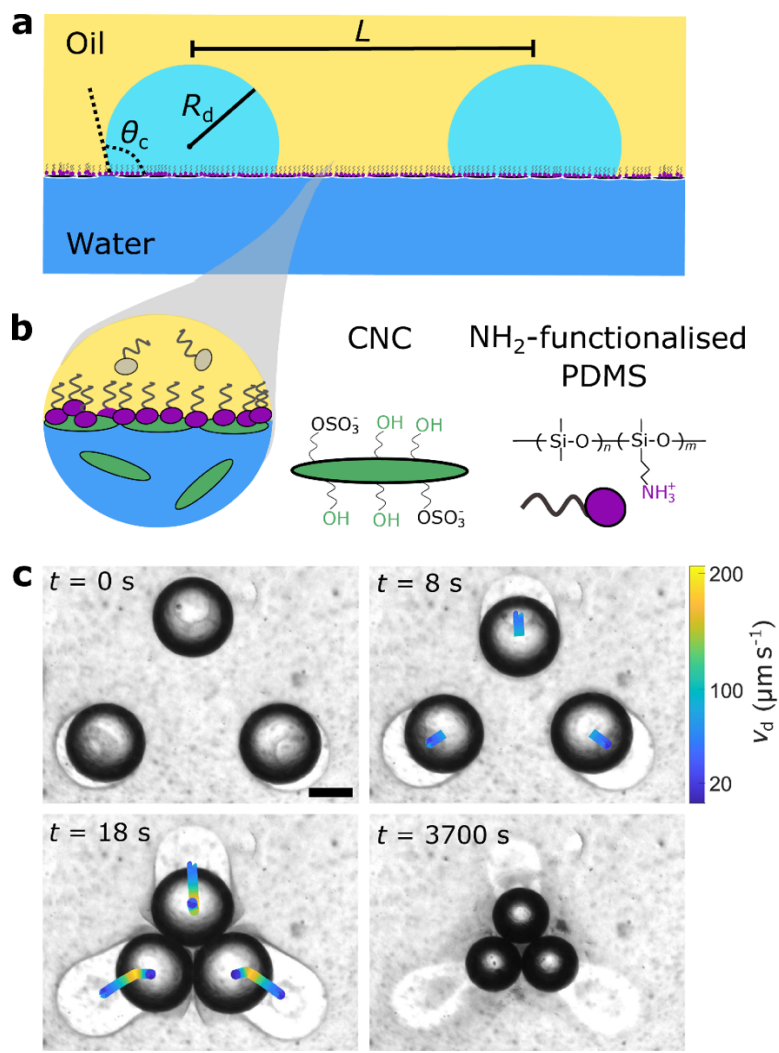

**Figure 1 | Placement and assembly of aqueous droplets at a solid-like oil-water interface. a)** Schematic of a water-ethanol droplet of radius $R_d$ and volume $V_d$ placed centre-to-centre distance $L$ from a second droplet. The droplets make a contact angle $\theta_c$ at a macroscopic oil-water interface at which nanoparticle surfactants are assembled. **b)** Zoom-in schematic showing cellulose nanocrystal surfactants (CNCS) at the oil-water interface composed of cellulose nanocrystals (CNC) and NH$_2$-functionalised polydimethylsiloxane (PDMS). $R_d$ is on the order of mm; CNC length and diameter are on the order of 500 and 10 nm, respectively. **c)** Placement ($t = 0$ s), mutual attraction ($t = 8$ s and $t = 18$ s), and survival ($t = 3700$ s) against coalescence of water-ethanol droplets at the solid-like oil-water interface. Overlaid lines show droplet trajectories colour-coded for speed $v_d$ as a function of position. At $t = 0$ s; $L = 4.3$ mm, $V_d = 2$ μL, $R_d = 0.95$ mm, and $\phi_{EtOH} = 0.5$. Scale bar, 1 mm.

We first design an oil-water interface capable of sustaining aqueous droplets for prolonged periods of time by assembling cellulose nanocrystal surfactants (CNCS) at the macroscopic interface between the two liquids (**Figure 1**a, Methods). To do so, we carefully pipette dodecane (oil-phase density, $\rho_o = 749$ kg m$^{-3}$) containing 5% w/w NH$_2$-functionalised PDMS-based statistical copolymer surfactant (poly[dimethylsiloxane-*co*-(3-aminopropyl)methylsiloxane]) on top of an aqueous glycine buffer (5 mM, pH 3) containing 1 mg mL$^{-1}$ cellulose nanocrystals (CNC) bearing sulfate half-ester surface groups (aqueous-phase density, $\rho_w = 998$ g m$^{-3}$). Within a broad pH range (pH 1 – 5), the surfactants and nanocrystals present opposing electrostatic charges (Fig. 1b), so that CNCS assemble at the liquid-liquid interface driven by the electrostatic attraction between the protonated NH$_2$-functionalised PDMS in the oil phase and the negatively charged CNC in the aqueous phase. This electrostatic interaction leads to the rapid formation of the CNCS and their irreversible adsorption to the oil-water interface, thus forming a dense assembly with significant interfacial shear elasticity and bending modulus.[26,27] We refer to this interfacially assembled material as a 'solid-like oil-water interface'. This interface is pinned flat by the lip of a 3D-printed insert placed inside a transparent imaging chamber (Supplementary Figures S1 and S2, Methods). Due to its solid-like mechanical properties, droplets of water-ethanol mixtures (droplet volume, $V_d = 0.1 - 10$ μL) can be placed at this interface and sit there without coalescing with their underlying water phase (Methods, Supplementary Video S1). Droplet contact angles with the CNCS-coated oil-water interface were in the range $\theta_c = 110 - 130°$ (Supplementary Figure S3), comparable to values of $\theta_c$ for water droplets measured on bulk PDMS,[28] showing that the oil-side of the CNCS assembly is rather hydrophobic. To quantify the lifetime of the water-ethanol droplets at this interface as a function of ethanol volume fraction $\phi_{EtOH}$, we define a cut-off time of one hour above which we consider droplets to be 'long-lived'. This cut-off is long enough to assess resistance against coalescence within timescales relevant to the assembly of the droplets (Fig. 1c). We observe a broad range of $\phi_{EtOH}$ over which long-lived droplets can be placed at the oil-water interface ($\phi_{EtOH} = 0.05 - 0.8$, droplet density, $\rho_d = 991 - 845$ kg m$^{-3}$ (Ref. [29]), Supplementary Figure S4).

By investigating the placement of multiple aqueous droplets (Supplementary Video S2), it immediately emerges that they attract one another over centre-to-centre separations, $L$, up to ~5 mm (Fig. 1c, Supplementary Video S3). Initial droplet speeds at separations just below this distance are on the order of μm s$^{-1}$, reaching hundreds of μm s$^{-1}$ near droplet contact (Fig. 1c). All the above are indicative of a long-range attraction between the droplets that increases in strength as the droplets move closer to each other. Exchange of solvents between the droplets and the macroscopic aqueous and non-aqueous phases leads to the system's time evolution; for small numbers of droplets placed at the solid-like oil-water interface, this compositional ripening leads to reduction in volume of the droplets over hour-long timescales without coalescence either between the droplets or the underlying aqueous phase (Fig. 1c). These results demonstrate a promising, novel mechanism for building long-lived, aqueous cellular materials at liquid-liquid interfaces.

To harness this mechanism towards materials fabrication, it is important to determine and understand the origin of this attractive interaction between the droplets (**Figure 2**). Imaging the system from the side (Methods) allows us to visually estimate that a single droplet ($\phi_{EtOH} = 0.25$, $V_d = 2$ µL) vertically depresses the macroscopic oil-water interface by about 100 µm, with the interface returning to a planar configuration about 5 mm from the droplet centre (Fig. 2a). This deformation arises from a balance between buoyancy and the surface tension of the oil-water interface, $\gamma_{ow}$, the ratio of which is characterised by the Bond number, $Bo = \frac{\Delta \rho g R_d^2}{\gamma_{ow}}$, where $\Delta \rho = \rho_w - \rho_o = 249$ kg m$^{-3}$ and $g = 9.81$ m s$^{-2}$. The balance between these two stresses leads to a natural length scale, the capillary length, $\lambda_c = \sqrt{\frac{\gamma_{ow}}{\Delta \rho g}}$, which describes the characteristic decay distance of the strength of the capillary interactions. By numerically modelling interfacial deformation by a single droplet in the limit $Bo \ll 1$, we can estimate $\gamma_{ow}$ through visual comparison with side-view images, allowing us to constrain $\gamma_{ow}$ in the range $2 < \gamma_{ow} < 5$ mN m$^{-1}$ (Supplementary Derivation S1, Supplementary Figures S5 and S6), which is consistent with previous experimental measurements of the oil-water interfacial tension of a CNCS assembly immediately prior to solidification.[26] This estimated range of $\gamma_{ow}$ further allows us to constrain the capillary length to the range $1 < \lambda_c < 2$ mm, in good agreement with our visual estimate that the oil-water interface is effectively flat further than 5 mm away from a single droplet.

When two droplets are placed at the interface with a centre-to-centre distance $L \leq 5$ mm, the combination of the deformations of the interface by each droplet form a central depression that set both droplets in motion towards each other due to capillary interactions (Fig. 2a, Supplementary Video S4). We can quantify this attraction by studying the dynamics of a simple system of two identical droplets with varying radii ($R_d \in [0.6, 1.4]$ mm and, hence, volume $V_d \in [1, 6]$ µL) and compositions ($\phi_{EtOH} \in [0.2, 0.7]$) placed a controlled $L$ away from one another, and extracting their velocities using digital video microscopy. Measured droplet speeds $v_d$ are in the range $1 - 200$ µm s$^{-1}$, with $v_d$ increasing strongly with $V_d$ (and, hence, $R_d$) and showing inverse proportionality between $v_d$ and $L$ over a broad range of $L$ (Fig. 2b). To model these dynamics and gain further insight, we consider the balance of the capillary force $F_c$ between the two droplets, the Stokes drag $F_s$ between each droplet and the bulk oil, and the viscous drag $F_v$ between each droplet and the solid-like CNCS assembly (Supplementary Derivation S2).

In the limit $Bo \ll 1$, it can be shown that the attractive capillary force between two droplets varies as $F_c \sim R_d^6 K_1\left(\frac{L}{\lambda_c}\right)$, where $K_1$ is a modified Bessel function of the first kind and of order 1. Working in the limit of small droplet separations, $L \ll \lambda_c$, further allows us to obtain simple scaling relations using the approximation, $K_1\left(\frac{L}{\lambda_c}\right) \approx \frac{\lambda_c}{L}$, leading to $F_c \sim \frac{R_d^6}{L}$. The Stokes drag, $F_s \sim R_d v_d$, scales linearly with droplet radius and speed regardless of modifications for the presence of the solid-like CNCS assembly or the interfacial rheology of the droplet itself. Finally, because of the high contact angles for the droplets studied in this work ($\theta_c \approx 110 - 130°$, Supplementary Fig. S3), $F_v$ varies with the size of the contact area of the droplet with the interface, $F_v \sim R_d^2 v_d$. We note that this differs from the scaling identified in the specific regimes of very small or very large contact angles, which are not relevant for our system. The Reynolds number, $Re = \frac{\rho_d v_d R_d}{\mu_o}$ (oil-phase viscosity, $\mu_o = 1.34$ mPa s$^{-1}$), of the system is in the range $10^{-4} - 10^{-1}$, meaning droplet inertia can be neglected and hence, in this overdamped regime, the balance of the three previous forces gives $\mathbf{F_c} + \mathbf{F_v} + \mathbf{F_s} \approx 0$. From these expressions we can extract a scaling relation between droplet speed and separation of the form $v_d \propto L^{-1}$, having fixed the droplet size (i.e., $R_d$). This predicted inverse proportionality is in good agreement with the experimental data and holds well even outside the limit of small droplet separations (Fig. 2b). Deviations from our simple model are expected and found for large droplet separations, where frictional forces between the droplet and the CNCS assembly overcome capillary attraction and arrest droplet motion, and for very small droplet separations ($L \approx 2R_d$) where film drainage slows droplet attraction (Supplementary Figure S7). Deviations from the model can also be expected for larger droplets above the volumes used here for which the small Bo approximation is no longer valid.

The dependence of $F_c$, $F_s$, and $F_v$ on droplet size further allows us to relate the variation of the droplet speed to the droplet separation and size which, in the regime where drag between the droplet and the solid-like CNCS assembly dominates over Stokes drag, we show is given by (Supplementary Derivation S2):

$$v'_d(L^{-1}) \equiv \frac{dv_d}{d(L^{-1})} \sim R_d^4. \quad (1)$$

We note that, due to the presence of the CNCS, this differs from the $\sim R_d^5$ scaling predicted for a pristine liquid-fluid interface (Supplementary Derivation S2).[30] We can calculate $v'_d(L^{-1})$ from fits to the $v_d$ versus $L^{-1}$ data (Fig. 2b) and, by plotting it as a function of droplet size, we find good agreement with this predicted scaling of $\sim R_d^4$ (Fig. 2c). This allows us to describe our data using a single master curve (Fig. 2c) and confirms capillary interactions at the solid-like oil-water interface as the driving mechanism for the self-building pattern in Fig. 1c.

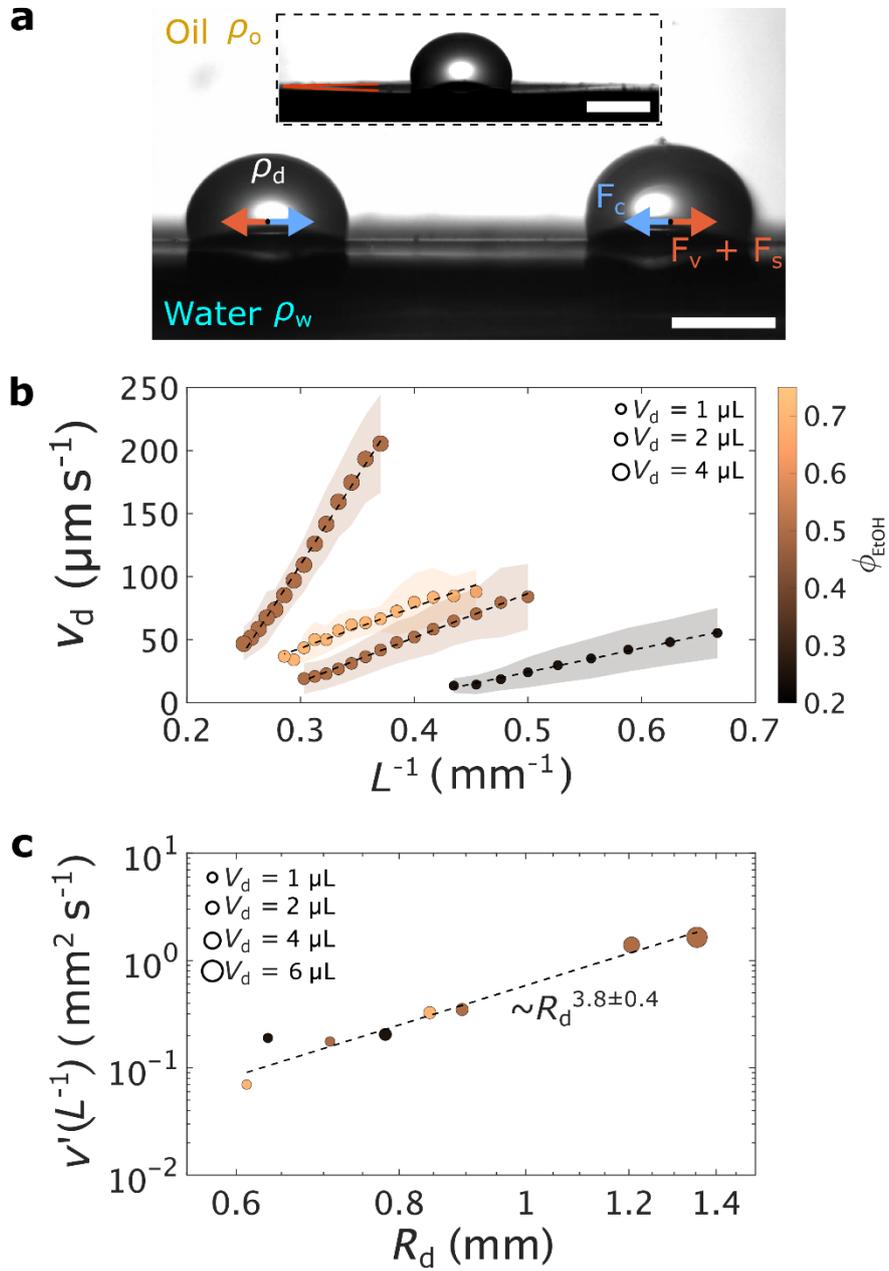

**Figure 2 | Scaling relations in the dynamics of droplets at a solid-like oil-water interface. a)** Deformation of the interface between oil (density, $\rho_o$) and water (density, $\rho_w$) by water-ethanol droplets (density, $\rho_d$). A solid-like cellulose nanocrystal surfactant assembly is present at the oil-water interface. Arrows show direction of capillary force $F_c$, Stokes drag $F_s$, and drag $F_v$ between the droplet and the oil-water interface. Inset: Deformation of the same interface by a single droplet. Red line shows deformation profile. Droplet radius $R_d = 0.8$ mm, droplet volume $V_d = 2$ µL, ethanol volume fraction $\phi_{EtOH} = 0.25$. Scale bars, 1 mm. **b)** Inverse proportionality of droplet speed $v_d$ and centre-to-centre separation $L$ due to attractive capillary interactions for droplets of different volumes $V_d$ (legend). Dashed lines show fits to the data (colour-coded for $\phi_{EtOH}$). Shaded region shows standard deviation in 3 measurements. **c)** Master curve of relation between droplet speed, size, and separation for the full range of parameter space studied in this work. The dashed line is a power law fit of $v'_d(L^{-1}) \equiv \frac{dv_d}{d(L^{-1})}$ versus $R_d$ in comparison with our theoretical prediction (Eq. 1). Each data point (colour-coded for $\phi_{EtOH}$) at different volumes $V_d$ (legend) consists of the average of at least 3 independent measurements. Standard error shown as error bars are smaller than the data points.

We can now harness our insight about the driving mechanism of droplet dynamics at the solid-like oil-water interface to build cellular materials. We can leverage both our ability to precisely print droplets at the interface and the capillary interactions between them to guide the self-assembly of cellular structures composed of large numbers of droplets (**Figure 3**). To do so, we need to quantify the lifetime of the droplets before coalescence with each other over a broad range of droplet compositions (Fig. 3a). Based on this data, we focus on equal composition droplets of $\phi_{EtOH} \leq 0.5$ to build long-lived multi-droplet structures. As well as directly depositing droplets by hand via a pipette, a bespoke droplet-deposition printer is integrated in our microscopy setup to allow precise dosing and positioning of droplets of varying size and composition at the oil-water interface (Methods). Given a droplet composition, we use this setup to control the initial locations of large numbers of droplets and investigate the structures that result from varying initial droplet position and size. Typical droplet deposition rates are $\sim 0.1 - 1$ s$^{-1}$, while the system dynamics that we have measured and discussed previously (Fig. 2) allow us to tune droplet assembly timescale from tens of seconds to minutes by varying initial droplet separations. We are thus able to engineer the assembly of multi-droplet structures from building blocks separated initially by mm-distances (Fig. 3b, Supplementary Video S5). Most simply, monodisperse systems assemble into hexagonal arrangements of droplets (Fig. 3c). Constructing a lattice from unit cells comprising equal numbers of large ($R_d = 0.75$ mm) and small droplets ($R_d = 0.35$ mm) of a diameter ratio of ≈0.47 leads to the formation of square lattices (Fig. 3d), in agreement with theoretical predictions for hard-sphere packing.[31] The structures are inherently dynamic during construction, undergoing significant plastic deformation in response to the addition of new droplets, including the annealing of lattice defects (Supplementary Video S6). This novel technique, which we term 'capillary-assisted printing', is to our knowledge the first method that combines the precise control over initial system conditions that printing enables with long-range capillary interactions between particles to produce a material that effectively builds itself.

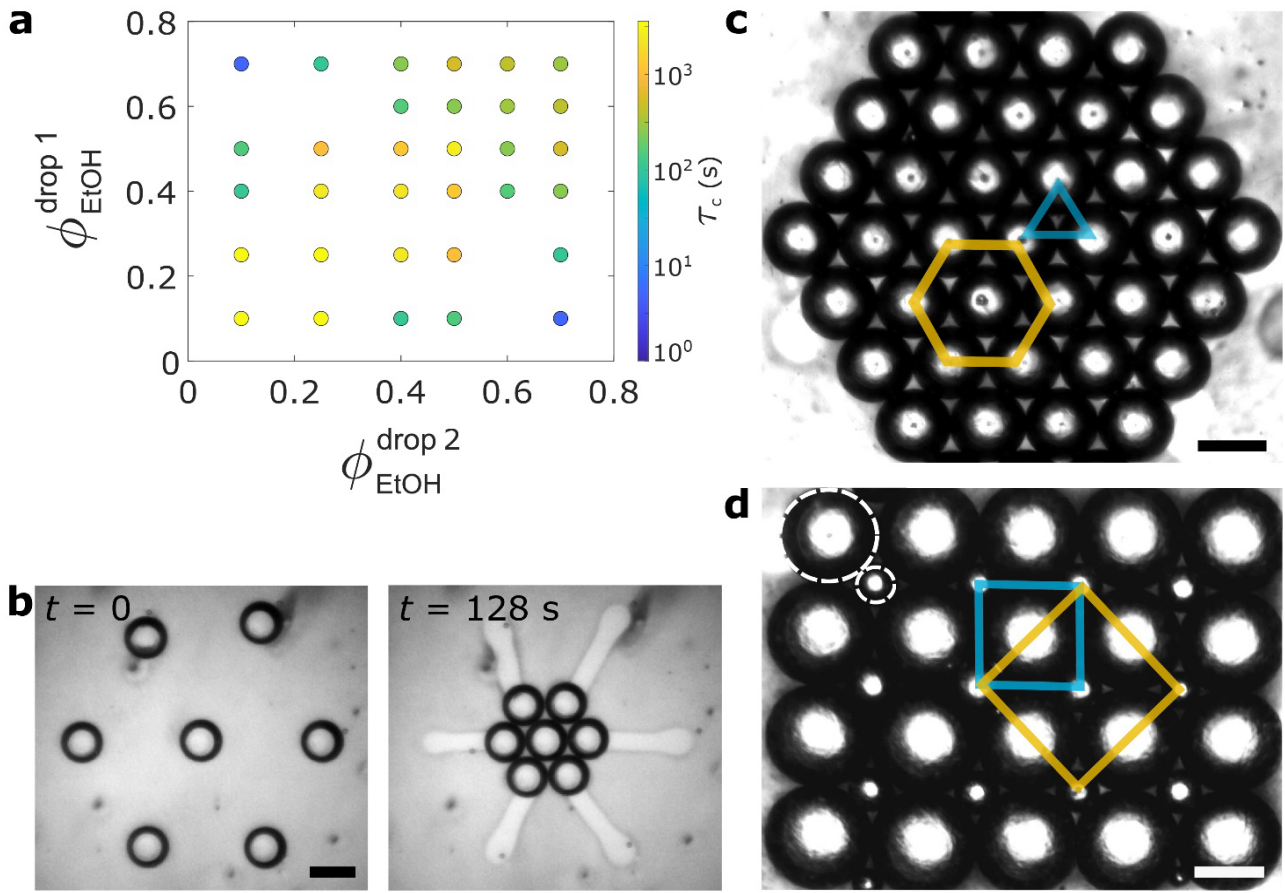

**Figure 3 | Printed, Self-Building Droplet Lattices**. **a**) Composition-dependence of the lifetime $\tau_c$ of two water-ethanol droplets against coalescence with each other. $V_d = 2$ μL. Empty circles show compositions for which single droplets could not be placed stably at the oil-water interface. **b**) Printed arrangements of monodisperse droplets assemble into a hexagonal heptamer ($\phi_{EtOH} = 0.25$, $V_d = 0.5$ μL) over the course of two minutes (time $t$ shown in seconds). **c**) As more monodisperse droplets join the assembly, hexagonal packing emerges over larger scales. Blue and yellow lines highlight lattice features ($\phi_{EtOH} = 0.25$, $V_d = 1$ μL). **d**) Square packing of bi-disperse water-ethanol droplets ($\phi_{EtOH} = 0.25$, initial volumes $V_d = 2$ μL and $V_d = 0.2$ μL). The dashed circles highlight the two droplet species. Scale bars, 1 mm.

Due to the compartmentalised liquid nature of droplets, our method allows us to print and assemble droplet-based structures with both compositional and structural diversity (**Figure 4**). First, we can embed droplets of heterogeneous composition within a self-building base layer (Fig. 4a-b) and then these rafts are compact enough to support another layer of droplets (Fig. 4c). In Fig. 4a-b, we show how we can print aqueous droplets containing dye into a matrix of undyed droplets. For the dye, we use rhodamine 6G which has both good solubility in water-ethanol droplets and has excitation and emission spectra that are well-characterised in water-ethanol solutions.[32] At $\phi_{EtOH} = 0.25$ the dye exhibits strong absorbance in the range 475 – 550 nm, giving dyed droplets a characteristic red colour under transmission imaging when illuminated with white light. In the system composition studied here, no dye exchange is observed between droplets, and the chemical and chromatic heterogeneity introduced by the dye can serve as a means to encode written texts or pictographic messages within non-dyed droplet assembles. We can therefore use this colouring as an additional dimension to explore the idea of embedding hidden information within the structure (Fig. 4a). Colour channel decomposition of images of the printed patterns allows the embedded patterns to be seen in the green channel, due to strong absorbance in this region, but not the red. Moreover, macrophotography of printed patterns from the side shows the rhodamine-containing droplets as green, in agreement with the fluorescence spectrum for the dye (Fig. 4b). The printing method we have developed makes it well versed to print droplet rafts containing hundreds of droplets. After assembly, these larger rafts can support an additional layer of dyed droplets placed on top of the base layer (Fig. 4c). As a simple proof-of-concept, we can place small numbers of rhodamine 6G-containing droplets to form a smiley face on top of a compositionally uniform base monolayer.

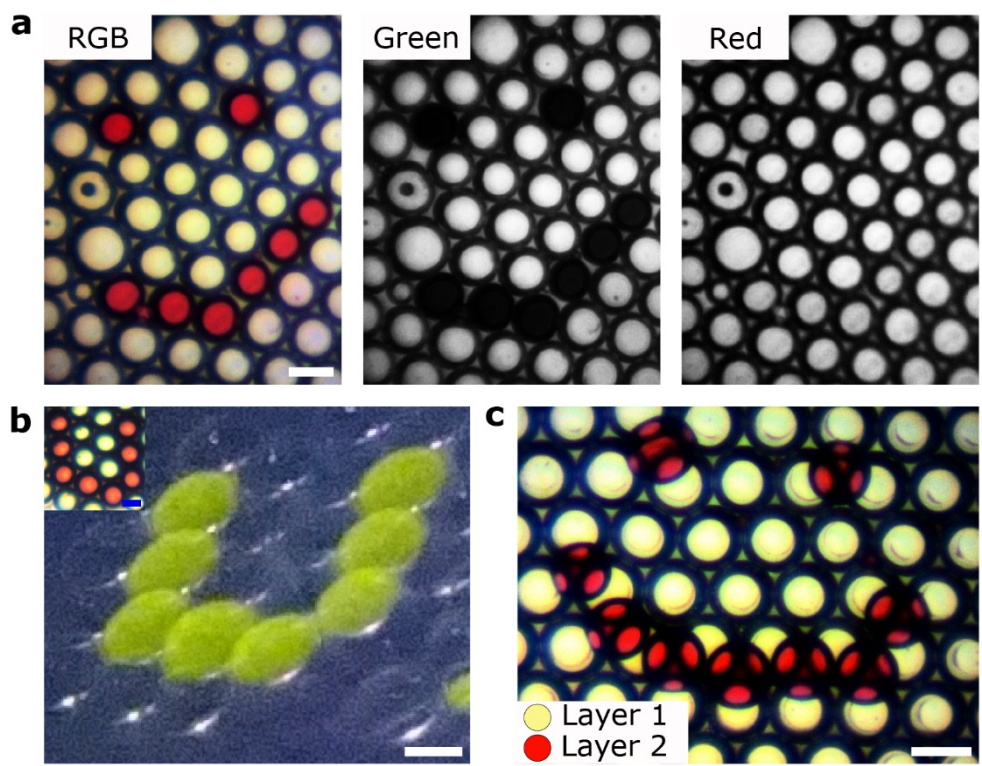

**Figure 4 | Compositional heterogeneity and multilayers in self-building droplet materials. a)** Colour decomposition of a hexagonal droplet lattice patterned with droplets that contain rhodamine 6G dye (red) to form a smiley face. **b)** Photograph taken at an angle to a printed droplet raft showing green fluorescence signal from droplets containing the dye (red in the transmitted light image, inset). **c)** Two-layer structure in which a smiley face pattern of dye-containing droplets is placed on top of a raft of aqueous droplets containing no dye. $\phi_{\text{EtOH}} = 0.25$, $V_{\text{d}} = 1$ μL. Scale bars, 1 mm.

So far in this work the CNCS assembly has only been used to extend the lifetime of water droplets placed at the oil-water interface and introduce capillary interactions. However, cellulose nanocrystal surfactants belong to the broad materials family of nanoparticle surfactants, which can be made from an enormous palette of components including MXenes,[33] polyoxometalates[34] and DNA.[35] This materials flexibility exists because the formation of nanoparticle surfactants is determined not by the composition of the nanoparticles themselves, but rather by the mutuality of the interactions between the functional groups of the polymer surfactant and the nanoparticle surface.[36] It is therefore possible to integrate nanoparticles into our CNCS assembly that can introduce response to external stimuli. For example, the introduction of gold nanoparticles can produce highly localised thermoplasmonic heating of the liquid phases (**Figure 5**). Heating of aqueous solutions by metal nanoparticles can readily generate temperature gradients of several K μm$^{-1}$ that in turn drive transport phenomena (including convection, Marangoni flows, and thermo-osmotic flows) that have been harnessed to manipulate microparticles, droplets, and even living cells.[37] Specifically, we add gold nanoparticles ('Au-NP', diameter = 11 – 25 nm) bearing tannic acid surface groups to the underlying aqueous phase of the interface. Under the conditions studied here (pH 3.0), the gold nanoparticles bear a negative surface charge and adsorb to the oil-water interface in the presence of the positively charged NH$_2$-functionalised PDMS. This adsorption leads to the formation of multi-component nanoparticle surfactant-covered interfaces containing both CNCS and gold nanoparticle surfactants

('Au-NPS', Fig. 5a). Strong absorbance of incident light at the localized surface plasmon resonance of the nanoparticles, which exhibits a broad peak with a maximum centred around 522 nm for the particles used here,[38] leads to strong, highly localised heating in the vicinity of the incident light and, near the oil-water interface, temperature-induced flows towards the heat source.[39] When a continuous wave laser of wavelength $\lambda = 532$ nm is focussed on the CNCS/Au-NPS assembly with a lens (power density $P \approx 160$ mW mm$^{-2}$, *Methods*), a single aqueous droplet ($\phi_{EtOH} = 0.4$, $V_d = 0.1$ μL) placed at this assembly is attracted towards the laser spot at speeds up to several hundred μm s$^{-1}$ (Fig. 5b, Supplementary Video S7) over distances in excess of several millimetres. We can further perform multi-step manipulations of this droplet across cm-distances over several minutes by shifting the position of the laser focus with respect to the droplet, allowing us to translate a droplet across two-dimensional trajectories (Fig. 5c). Beyond one droplet, this optofluidic 'handle' allows us to attract multiple droplets to a single pre-determined point, overcoming the capillary interaction to guide their assembly into structures (Fig. 5d, Supplementary Video S8). Moreover, at droplet contact, the capillary attraction between the droplets is sufficiently strong to hold multi-droplet constructs together in the presence of the thermally driven flows, so that we can coherently translate multi-droplet structures using this thermoplasmonic effect (Fig. 5e). This is the first example of plasmon-assisted optofluidic construction of droplet-based structures at liquid-liquid interfaces and opens a new pathway for building and manipulating liquid materials.

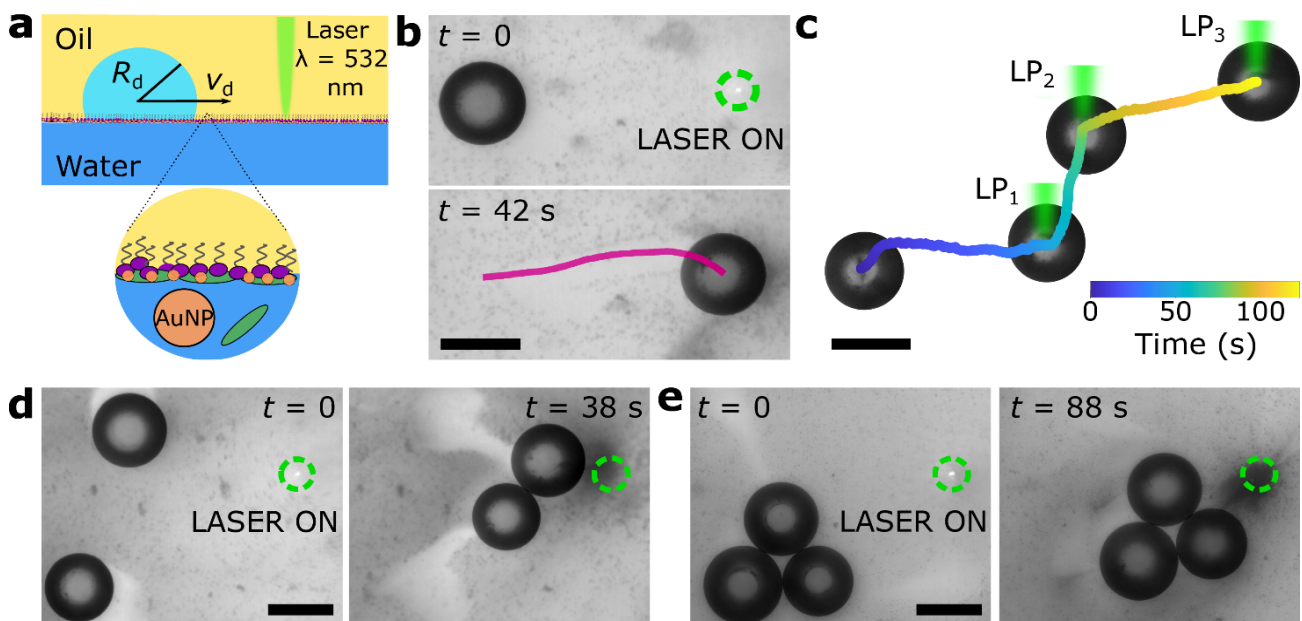

**Figure 5 | Thermoplasmonic manipulation of droplets at a solid-like, gold-nanoparticle-functionalised, oil-water interface. a)** Schematic showing a droplet of radius $R_d$ placed on an oil-water interface at which a solid-like layer of cellulose nanocrystal surfactants (CNCS) and gold nanoparticle surfactants (Au-NPS) are adsorbed. A laser of wavelength $\lambda = 532$ nm is focussed with a lens at the assembly (power density $P \approx 160$ mW mm$^{-2}$), which drives motion of the droplet towards the laser at a velocity $v_d$. **b)** Motion of a droplet towards a laser focussed at the solid-like oil-water interface. The position of the laser is shown by the green dashed circle, the purple line shows the droplet trajectory over 42 s. **c)** Multi-step, controlled translation of a single droplet using plasmonic heating. LP$_1$, LP$_2$ and LP$_3$ denote the first, second and third points to which the laser was moved, respectively. Coloured line shows droplet position as a function of time (see colour legend). Background removed for clarity. **d)** Optofluidic assembly of a droplet dimer and **e)** optofluidic translation of a droplet raft, here a trimer. $V_d = 0.1$ μL, $\phi_{EtOH} = 0.4$. Scale bars, 0.5 mm.

In summary, we have developed a method, which we term 'capillary printing', to print, assemble, and manipulate structures of long-lived aqueous droplets at liquid-liquid interfaces. This was achieved by placing the droplets onto a macroscopic oil-water interface at which a solid-like layer of nanoparticle surfactants was assembled. The deformation of this solid-like oil-water interface by the droplets leads to long-range attractions between them with system dynamics that are well described by a classical theory of capillary forces that we modify to account for the drag between the droplet and the solid-like oil-water interface. This leads to novel scaling behaviour, allowing us to describe our data using a single master curve. A droplet printer was developed that could precisely define the initial positions of the droplets, allowing us to fabricate highly plastic self-building droplet lattices of heterogeneous composition. Finally, gold nanoparticles were integrated into the nanoparticle surfactant assembly, giving us an optofluidic handle with which to manipulate the system. Strongly localised thermoplasmonic heating caused by focussing a laser at the oil-water interface led to temperature-induced flows and hence fine-tuned control over droplet position manipulation. Future work must focus on generalising the formulation of the system to incorporate biologically derived, living, and active materials, and on harnessing the nanoparticle surfactant assembly to introduce other forms of control by external stimuli, e.g., magnetic response or catalytic functionality. Our results demonstrate a new pathway for fabricating and manipulating complex, cellular materials made entirely from liquids with potential applications in reconfigurable microfluidics, adaptive optics, and systems chemistry.[40,41]

**Acknowledgments**
The authors would like to thank Joe Keddie for insightful suggestions. JF was supported by a Ramsay Memorial Fellowship. AT acknowledges Valentino Barbieri for fruitful discussions on gold nanoparticles and Sandrine Heijnen for help with the laser setup. RM and GV gratefully acknowledge the Engineering and Physical Sciences Research Council for supporting this work [grant numbers EP/W005875/1]. GV and MB acknowledge the Engineering and Physical Sciences Research Council for supporting this work and for providing AT with a research scholarship [grant number EP/R513143/1].

**Author Contributions**
Author contributions are defined based on the CRediT (Contributor Roles Taxonomy). **AT**: Conceptualisation, Methodology, Software, Validation, Formal Analysis, Investigation, Data Curation, Writing (Original Draft), Visualisation. **RM**: Conceptualisation, Validation, Methodology, Writing (Review and Editing). **MB**: Conceptualisation, Writing (Review and Editing), Methodology, Supervision. **GV**: Conceptualisation, Data Curation, Methodology, Resources, Writing (Review and Editing), Supervision, Funding Acquisition. **JF**: Conceptualisation, Methodology, Investigation, Software, Writing (Original Draft).

**Conflicts of Interest**
There are no conflicts of interest to declare.

**Data Availability**
All data available upon request to j.forth@liverpool.ac.uk.

**Code Availability**
A *jupyter notebook* containing all numerical modelling code used in this work can be found at https://github.com/joeforth/droplet-model.

# Methods

## Materials

Cellulose nanocrystals (length: 300 - 900 nm, diameter: 10 - 20 nm) were purchased from Nanografi Nano Technology. Glycine (≥ 99 %), poly[dimethylsiloxane-co-(3-aminopropyl)methylsiloxane]), n-Dodecane (≥99 %), rhodamine 6G (~95 %), and tannic acid-functionalised gold nanoparticles (diameter: 11 - 25 nm; $5.3 – 6.58 \times 10^{12}$ particles mL$^{-1}$) were purchased from Sigma-Aldrich / Merck. Ethanol (99%+) was purchased from Fisher Scientific. All materials were used without further purification. All water used to prepare aqueous solutions was obtained by reverse osmosis (resistivity > 18 MΩ cm, Elga Purelab).

## Preparation of liquid phases for the oil-water interface

The aqueous phase was prepared by dispersing 1 mg mL$^{-1}$ cellulose nanocrystals in a 5 mM glycine solution in water at pH 3. pH was adjusted using 1M HCl. The resulting suspension was sonicated for 15 minutes. The oil phase was prepared by dispersing 5% w/w poly[dimethylsiloxane-co-(3-aminopropyl)methylsiloxane] in n-dodecane. For the experiments in Fig. 5, aqueous suspensions of gold nanoparticles and CNC were mixed at a volume ratio of 0.4:0.6 respectively, yielding a CNC concentration of 0.6 mg mL$^{-1}$. All batch aqueous solutions containing cellulose nanocrystals and/or gold nanoparticles were stored in the refrigerator (~ 4 °C) and were equilibrated to room temperature and sonicated briefly before experiments. All other solutions without nanomaterials were sealed with parafilm and stored in an environmentally controlled room (temperature: 20 ± 1 °C, relative humidity: 45 ± 5 %).

## Formation of CNCS assembly at the oil-water interface

Macroscopic oil-water interfaces were prepared by pipetting the aqueous phase into either a cuboidal quartz glass cuvette (Krüss GmbH, inner dimensions: 20 × 20 × 20 mm$^3$) or, if a larger surface area was required, into an optically clear flat-bottom Petri dish of 90 mm inner diameter and depth of 20 mm (Supplementary Fig. S1). The frosted glass base of the cuvette was rendered transparent by using a drop of immersion oil (refractive index, 1.51) to optically match the glass with a microscope slide placed underneath the cuvette. Both sample containers were fitted with a 3D-printed polylactic acid (PLA) trough that pinned the oil-water interface flat (Supplementary Figs. S1 and S2). The oil phase was then carefully layered on top with a 2:1 oil-water volume ratio. The height of the aqueous phase was kept at 10 mm and that of the oil phase at 4 mm in all experiments. The sample was then allowed to rest for 120 minutes while the CNCS layer assembled, after which aqueous droplets could be placed at the oil-water interface.

## Droplet preparation, deposition and printing

Ethanol volume fraction, $\phi_{EtOH}$, was calculated as $\phi_{EtOH} = \frac{V_{EtOH}}{V_{EtOH}+V_{H_2O}}$, where $V_{EtOH}$ and $V_{H_2O}$ were the initial volumes of ethanol and water, respectively, that were added to the mixture. For droplets with rhodamine 6G, the dye was directly dissolved into the water-ethanol solutions at a concentration of approximately 1 mg mL$^{-1}$. Droplets were dispensed via either a mechanical pipette (Eppendorf) or a single-channel syringe pump (Aladdin 1002X, WPI) attached to a needle (14-gauge, 1" long, Fisnar) for automated deposition. The needle was clamped to a custom-built column next to the microscope setup via a motorised stage (Thorlabs, MT1/M-Z9) to allow for its vertical displacement over the sample plane. Relative horizontal displacements of the needle were obtained by moving the sample placed on a motorised two-dimensional stage (Newport, M-406). MATLAB code was developed in-house to program the syringe pump and the motorised stages, enabling automatic dispensing of droplets of different sizes at specific locations within the sample.

**Optical setup and microscopy**
Droplets at the oil-water interface were observed using an inverted microscope built in-house. A monochrome light emitting diode (LED) source (590 nm, M595L4, Thorlabs) illuminated the sample, which was placed on a motorised two-dimensional stage (Newport, M-406) and imaged onto a complementary metal-oxide semiconductor (CMOS) camera through a two-lens system. The setup also allowed for side-view using a different LED source (470 nm, M470L5-C2, Thorlabs). Colour viewing was achieved by switching to a colour camera (Thorlabs, DCC1645C) and a white LED (Thorlabs, MNWHL4). Droplet contact angles were measured using FIJI.[42] Droplets were tracked by digital video microscopy with a MATLAB code developed in-house starting from the images projected on the CMOS camera with an acquisition rate of 10 frames per second. Briefly, the algorithm performed background subtraction, image binarisation, and noise removal using the in-built *bwareopen* function and extraction of the droplet's centroid coordinates and radius using the *regionprops* function.

**Thermoplasmonic manipulation**
Plasmonic heating experiments were performed by illuminating the samples with a 532 nm continuous wave (CW) laser (Gem, Laser Quantum). This required a bespoke setup to perform simultaneous imaging of the system under focussed laser illumination. The laser beam was guided through a periscope composed of two sets of mirrors and then passed through a single lens with a focal length of 200 mm. The beam was then directed towards the sample with a 605 nm long-pass dichroic mirror (Thorlabs, DMLP605), and ultimately focused onto the oil-water interface of the sample. The diameter of the beam at the sample was consistently maintained at ~ 0.14 mm delivering a power density $P \approx 160$ mW mm$^{-2}$. The sample was mounted on an inverted optical microscope (Leica, DMI 4000B) equipped with a two-dimensional motorised stage (Microdrive, MCL), allowing for precise control over the sample's position relative to the stationary laser beam. Illumination for imaging was provided by a white LED source (Thorlabs, MWWHLP1) located above the dichroic mirror. To enable a broad field of view, we used an 80-mm objective lens for imaging the sample with a monochrome CMOS camera (Thorlabs, DCC1545M) from below the sample plane. A long-pass filter (590 nm, FGL590nm, Thorlabs) was placed in front of the camera to filter out the green laser light, ensuring camera protection and image clarity.

# Supporting Information for Capillary-Assisted Printing of Droplets at a Solid-Like Liquid-Liquid Interface

Anshu Thapa[1], Robert Malinowski[1], Matthew O. Blunt[1], Giorgio Volpe[1*], Joe Forth[1,2,3*]

[1]Department of Chemistry, University College London, 20 Gordon Street, London, WC1H 0AJ, UK
[2]Department of Chemistry, University of Liverpool, Crown Street, Liverpool, L69 7ZD, UK
[3]Department of Physics, University of Liverpool, Oxford Street, Liverpool, L69 7ZE, UK

*Email: j.forth@liverpool.ac.uk; g.volpe@ucl.ac.uk

This file includes:

Supplementary Figures S1 – S7
Supplementary Derivations S1 and S2
Supplementary References
Captions for Supplementary Videos S1 – S8

**Supplementary Figures**

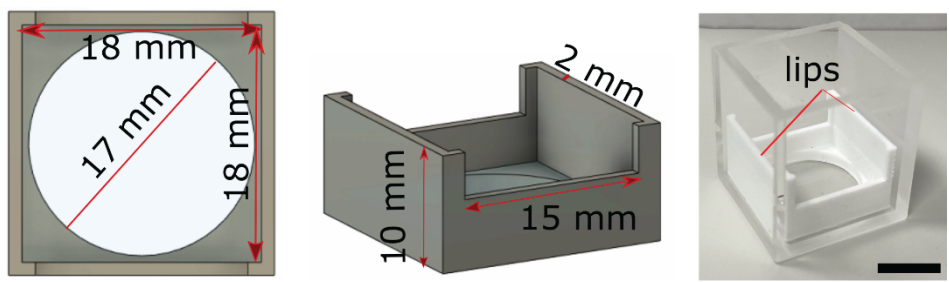

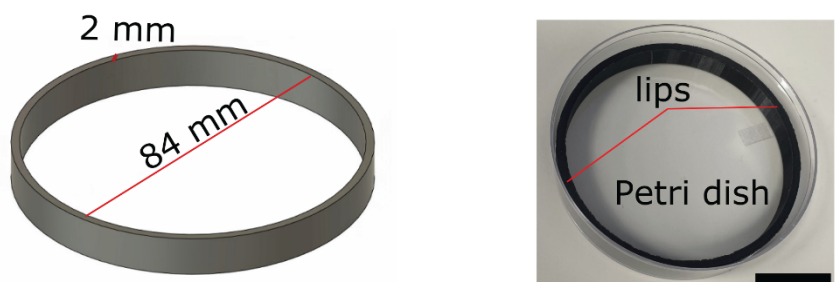

**Supplementary Figure S1 | Sample containers and 3D-printed inserts used to produce flat oil-water interfaces. a)** Custom-made 3D-printed polylactic acid (PLA) trough for quartz cuvette with lips to pin the three-phase contact line. The CAD drawings show top and side view of the trough with true dimensions. Scale bar, 1 cm. **b)** Custom-made 3D-printed PLA trough with lips for Petri dish. Scale bar, 2 cm.

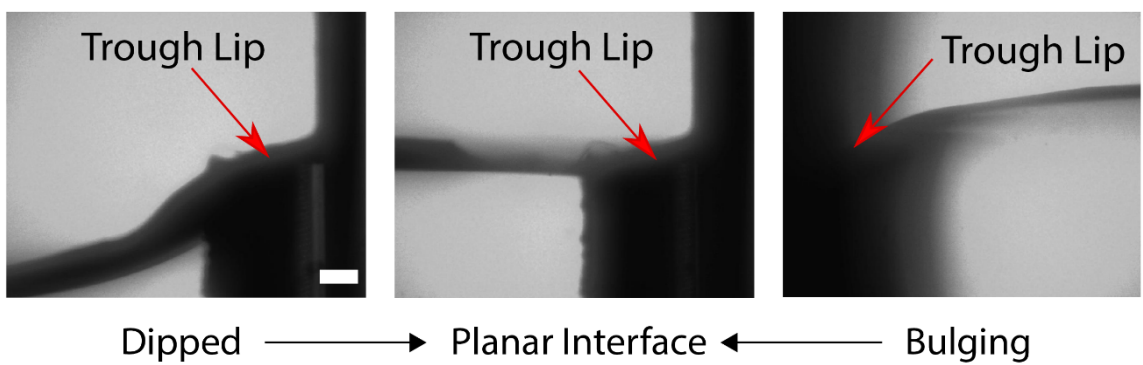

**Supplementary Figure S2 | Pinning of the oil-water interface into a flat configuration.** Flattening of the interface at the trough lip. The series of images shows how the level of the interface can be adjusted by varying the volume of the aqueous phase to produce a flat interface (centre image), using the trough lip to pin the three-phase contact line. Scale bar, 1 mm.

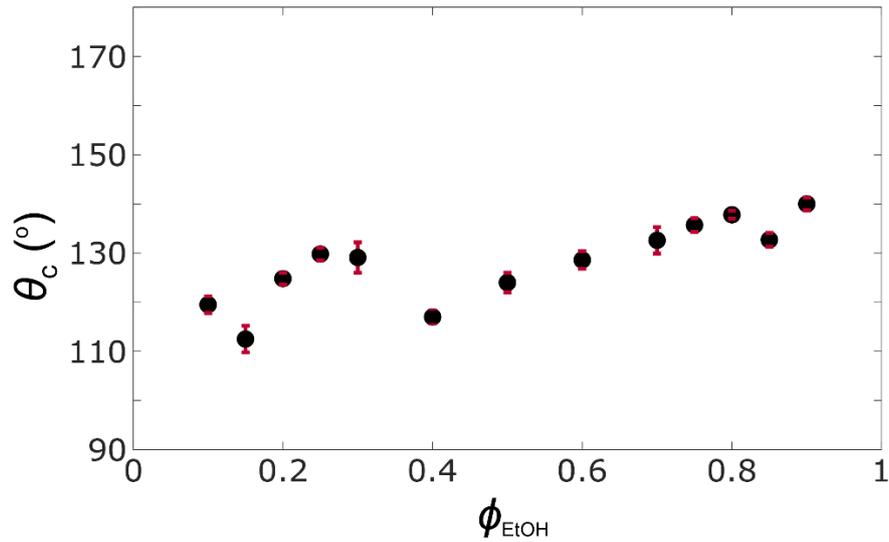

**Supplementary Figure S3 | Contact angle of water-ethanol droplets at a solid-like oil-water interface.** Contact angles measured for 2-µL stationary droplets. Each data point is the average of three or more experiments and the error bars represent the standard error. Measurements were taken within the first 60 seconds after droplet contact with the solid-like oil-water interface.

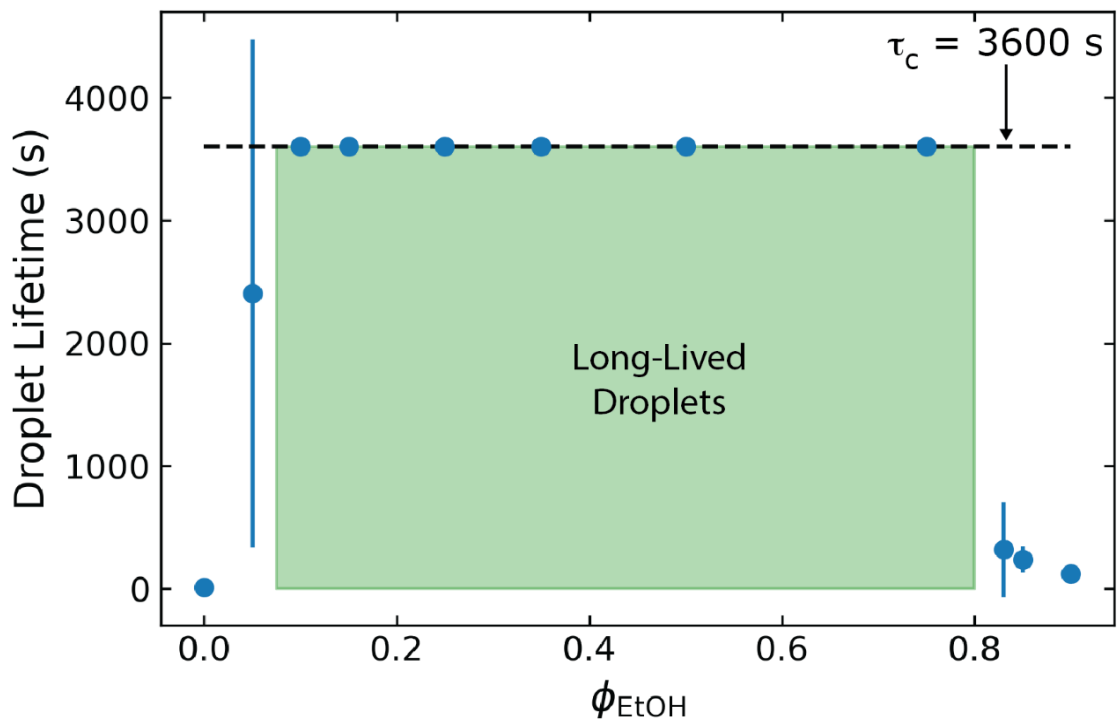

**Supplementary Figure S4 | Droplet lifetime against coalescence with the underlying aqueous phase of the solid-like oil-water interface.** Droplet lifetime as a function of ethanol volume fraction $\phi_{EtOH}$ of water-ethanol droplets against coalescence with the underlying aqueous phase of the solid-like oil-water interface. The green region shows the range of compositions over which droplet lifetime exceeds the experimental cut-off timescale ($\tau_c$ = 60 min). Each data point is the average of three or more experiments and the error bars represent the standard deviation.

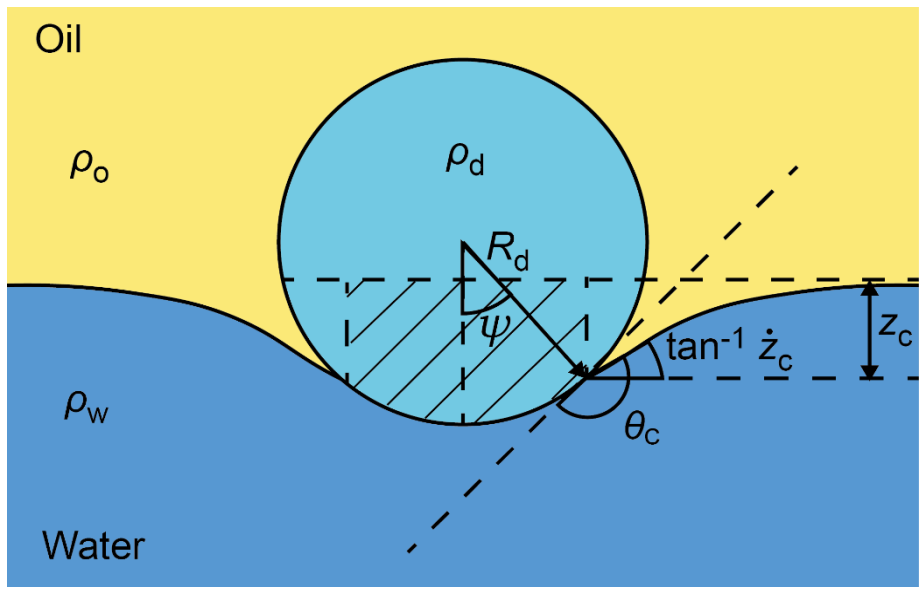

**Supplementary Figure S5 | Deformation of an oil-water interface by a droplet.** Schematic defining relevant quantities in our model of an aqueous droplet (density $\rho_d$) at an interface between oil (density $\rho_o$) and water (density $\rho_w$) with contact angle $\theta_c$. $R_d$: droplet radius; $\psi$: angle subtended between the droplet base and the three-phase contact line; $z_c$: depth of the interface deformation to the three=phase contact line.

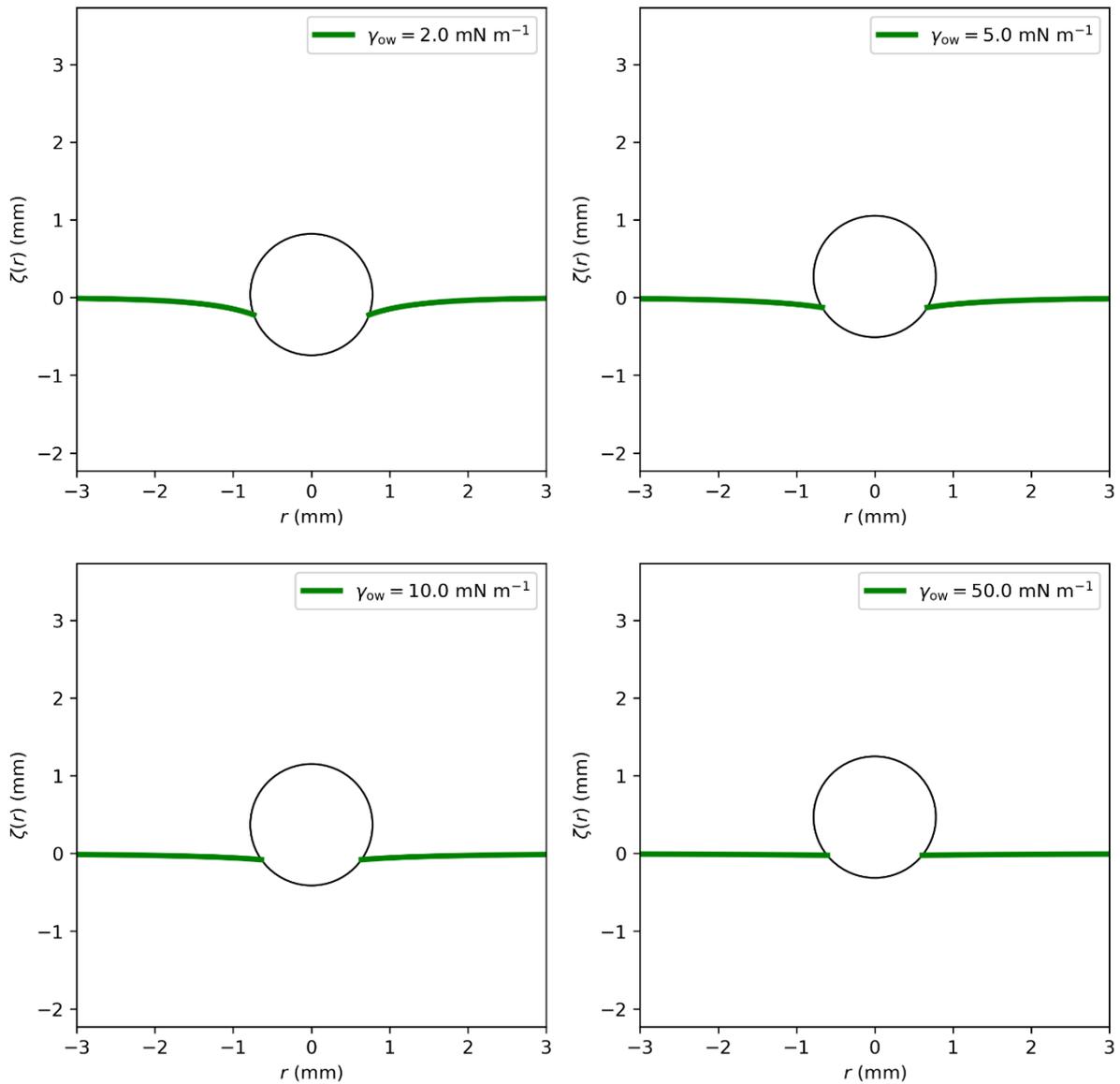

**Supplementary Figure S6 | Calculated deformation of the oil-water interface by a spherical object and its variation with surface tension.** Vertical displacement of the oil-water interface $\zeta(r)$ plotted as a function of distance from droplet centre, $r$, for varying values of oil-water surface tension, $\gamma_{ow}$. Droplet parameters and composition identical to the droplets shown in Fig. 2a: $\phi_{EtOH}$ = 0.25, $V$ = 2 μL, $\theta_c$ = 130°. $\gamma_{ow}$ values shown in the legend of each plot. Method of calculation derived in Supplementary Derivation S1.

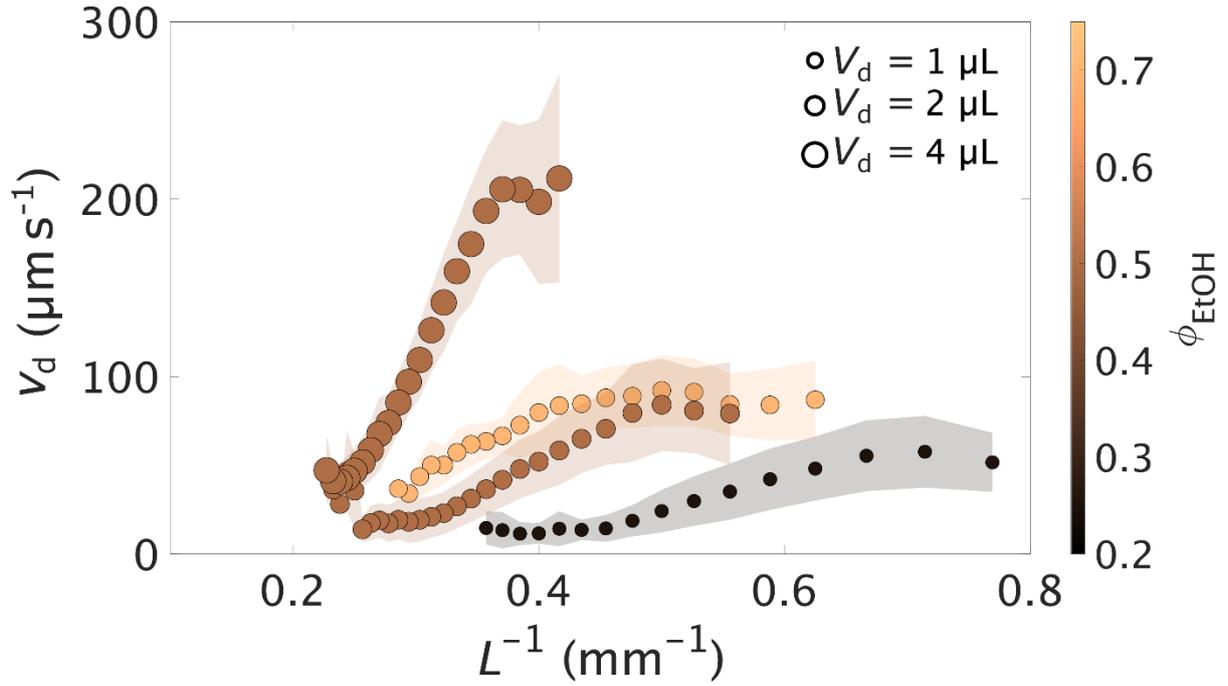

**Supplementary Figure S7 | Droplet speeds versus inverse centre-to-centre separation over the full range of droplet separations studied.** Extended data for Fig. 2b.

## Supplementary Derivations

**Supplementary Derivation S1: Calculating Interfacial Deformation by the Droplet**

To model our system, we adapt the framework of Nicolson (and later Chan *et al.* and Vella and Mahadevan),[1–3] which considers a spherical droplet of radius $R_d$ and density $\rho_d$ sat at the interface between two liquids, oil and water, with densities $\rho_o$ and $\rho_w$ respectively (and $\Delta\rho = \rho_w - \rho_o$) and with interfacial tension $\gamma_{ow}$ (Supplementary Figure S5). This model is derived in the regime where the ratio of buoyancy-to-surface-tension-derived stresses (i.e., the Bond number) is small, $\mathrm{Bo} = \frac{\Delta\rho g R_d^2}{\gamma_{ow}} \ll 1$, and the deformation of the droplet due to both buoyancy and interactions with the interface is neglected, so that the droplet can be treated as spherical. We note the trigonometric relationship between the contact angle of the droplet at the oil-water interface, $\theta_c$, the angle subtended between the base of the droplet and the three-phase contact line, $\psi$, and the slope of the oil-water interface at the three-phase contact line, $\dot{z}_c$:

$$\psi = \pi - \theta_c + \tan^{-1} \dot{z}_c \qquad (1)$$

We can calculate the deformation of the interface from the balance between buoyancy, which deforms the interface, and surface tension, which acts as a restoring force.[1] At any point on the oil-water interface with location Cartesian coordinates $\zeta(x_1, x_2)$, its two principal curvatures are given by:

$$\kappa_i = \frac{\partial^2 \zeta}{\partial x_i^2} \left(1 + \left(\frac{\partial \zeta}{\partial x_i}\right)^2\right)^{-3/2} \tag{2}$$

with $i = 1, 2$. For small curvatures, this can be simplified to $\kappa_i \approx \frac{\partial^2 \zeta}{\partial x_i^2}$ and so $\kappa_1 + \kappa_2 \approx \nabla^2 \zeta$. The pressure jump across the curved oil-water interface with surface tension $\gamma_{ow}$ is therefore given by:

$$\Delta P = \gamma_{ow}(\kappa_1 + \kappa_2) = \gamma_{ow} \nabla^2 \zeta \tag{3}$$

This jump in pressure must conform to the law of hydrostatic pressure:

$$\Delta P = \Delta \rho g \zeta \tag{4}$$

Equating the two terms gives:

$$\gamma_{ow} \nabla^2 \zeta = \Delta \rho g \zeta \tag{5}$$

Switching to cylindrical coordinates, in which the vertical position of the oil-water interface is described by $\zeta(r, \theta)$, and, assuming polar symmetry of the spherical droplet, i.e., $\nabla^2 = \frac{d^2}{dr^2} + \frac{1}{r}\frac{d}{dr}$, we can then write:

$$\frac{d^2 \zeta}{dr^2} + \frac{1}{r}\frac{d\zeta}{dr} - \frac{1}{\lambda_c^2}\zeta = 0 \tag{6}$$

Where $\lambda_c = \sqrt{\frac{\gamma_{ow}}{\Delta \rho g}}$ is the capillary length. The above differential equation has solutions of the form:

$$\zeta(r) = A K_0\left(\frac{r}{\lambda_c}\right) \tag{7}$$

where $K_n(x)$ is a modified Bessel function of order $n$. To find $A$, we need to impose boundary conditions. Far from the droplet, $\zeta \to 0$ as $r \to \infty$. Near the droplet, we use the fact that:

$$\frac{dK_0(x)}{dx} = -K_1(x) \tag{8}$$

and the definition of the gradient of the interface at the three-phase contact line

$$\dot{z}_c = \frac{d\zeta}{dr}\bigg|_{r_c} \tag{9}$$

which gives

$$\zeta(r) = \dot{z}_c \lambda_c \frac{K_0(\lambda r)}{K_1(\lambda r_c)} \tag{10}$$

In order to calculate $\zeta(r)$, one must then obtain $\dot{z}_c$, which Chan *et al.* have shown can be obtained from the balance of:[2]

1. The weight of the droplet:

$$-\frac{4\pi}{3} R_d^3 \rho_d g \tag{11}$$

2. The vertical component of the surface tension force:

$$2\gamma_{ow} \sin(\tan^{-1} \dot{z}_c) = 2\pi R_d \gamma_{ow} \frac{\dot{z}_c \sin\psi}{(1+\dot{z}_c^2)^{1/2}} \tag{12}$$

3. And upthrust due to displacement of the liquid phases:

$$\frac{4\pi}{3} R_d^3 \rho_o g - 2\pi R_d^3 \Delta\rho g \left[\frac{1}{2}\left(\frac{z_c}{R_d} + \cos\psi\right)\sin^2\psi - \frac{1}{3}(1-\cos^3\psi)\right] \tag{13}$$

The expression for upthrust can be simplified (and $z_c$ removed) by noting that in the limit of Bo $\ll$ 1, $\frac{z_c}{R_d} \to 0$ (i.e., the interface is flat when Bo = 0), so that terms containing $\frac{z_c}{R_d}$ can be neglected, giving:[2]

$$\dot{z}_c \sin\psi = \text{Bo}\left(\frac{2}{3}D - \frac{1}{3} - \frac{1}{2}\cos\theta_c + \frac{1}{6}\cos^3\theta_c\right) = \text{Bo}\,\Sigma \tag{14}$$

where $D = \frac{\rho_d - \rho_o}{\Delta\rho}$, and $\Sigma = \frac{2}{3}D - \frac{1}{3} - \frac{1}{2}\cos\theta_c + \frac{1}{6}\cos^3\theta_c$ acts as a dimensionless effective weight of the droplet.[3] We can explicitly calculate the deformation profile of the interface due to the droplet by combining Eqs. 1 and 14 to get an expression in terms of $\psi$:

$$\sin\psi \tan(\psi + \theta_c - \pi) - \alpha = 0 \tag{15}$$

where $\alpha = \text{Bo}\Sigma$. Equation 15 is then solved numerically (see *Code Availability* in main text) to find $\psi \in (0, \pi)$. Both $r_c = R_d \sin\psi$ and $\dot{z}_c = \tan(\psi + \theta_c - \pi)$ can now be calculated, allowing the deformation of the interface $\zeta(r)$ to be calculated from Eq. 10 (Supplementary Fig. S6).

**Supplementary Derivation S2: Equation of Motion for the Droplet**
In the overdamped regime (Re $< 10^{-1}$ in our system), the droplet dynamics are well captured by the balance of four different forces:

$$\mathbf{F}_v + \mathbf{F}_s + \mathbf{F}_f + \mathbf{F}_c \approx 0 \tag{16}$$

where $\mathbf{F}_v$ is the viscous drag force between the droplet and the solid-like oil-water interface containing the CNCS layer, $\mathbf{F}_s$ is the Stokes drag between the droplet and the external oil phase, $\mathbf{F}_f$ is the resistance to motion due to pinning of the three-phase contact line, and $\mathbf{F}_c$ is the capillary force between two droplets due the deformation of the oil-water interface caused by the same droplets. As we consider only system dynamics here, we neglect $\mathbf{F}_f$ in the following derivation, which has been shown to have negligible speed-dependence at the droplet speeds studied in this work ($v_d < 300$ μm s$^{-1}$).[4]

**$F_c$ – Capillary Interaction**

To calculate $F_c$, we can apply the observation of Chan *et al.* that the interaction energy between two droplets is the product of the effective weight of the first droplet and its vertical displacement due to the presence of the second droplet a distance $L$ away.[2] For Bo $\ll 1$, Nicolson has argued, in agreement with later numerical work, that $\zeta(r)$ for two particles is simply equal to the sum of the deformation from each individual particle.[1,5] After Chan *et al.*, the interaction potential $E(L)$ between two identical spherical objects can thus be written as:[2]

$$E(L) = -\frac{2\pi\gamma_{ow}R_d^6 \Sigma^2}{\lambda_c^4} K_0\left(\frac{L}{\lambda_c}\right) \qquad (17)$$

And the magnitude of the capillary force $F_c$ exerted by one droplet upon the other is given by:

$$F_c = -\frac{\partial E}{\partial L} = -\frac{2\pi\gamma_{ow}R_d^6 \Sigma^2}{\lambda_c^5} K_1\left(\frac{L}{\lambda_c}\right) \qquad (18)$$

using the relation $\frac{dK_0(x)}{dx} = -K_1(x)$. Working in the limit $\frac{L}{\lambda_c} \ll 1$, allows us to extract simple scaling relations from Eq. 17 by applying the approximation, $K_1\left(\frac{L}{\lambda_c}\right) \approx \frac{\lambda_c}{L(t)}$ giving:

$$F_c = -\frac{2\pi\gamma_{ow}\Sigma^2 R_d^6}{\lambda_c^4}\frac{1}{L(t)} = -\frac{2\pi\Sigma^2 \Delta\rho^2 g^2 R_d^6}{\gamma_{ow}}\frac{1}{L(t)} \qquad (19)$$

From which we note the scaling:

$$F_c \sim \frac{R_d^6}{L}. \qquad (20)$$

**$F_s$ – Modified Stokes Drag Near a Solid Wall**

The drag between the droplet and the bulk oil phase can be estimated from the Faxén correction to the Stokes drag for a spherical object moving with a velocity $v_d$ in a bulk liquid of viscosity $\mu$ modified by a coefficient $\beta$ to describe the reduction in drag due to the proximity of the droplet to the solid-like liquid-liquid interface, the magnitude of which is given by:[6,7]

$$F_s = 6\pi\mu R_d \beta v_d \qquad (21)$$

In this case, $\beta = \frac{1}{1-\frac{9}{16}\left(\frac{R_d}{s}\right)+\frac{1}{8}\left(\frac{R_d}{s}\right)^3}$, where $s$ is the distance of the particle's centre to the interface. For the case of a spherical particle in contact with the surface, $\frac{R_d}{s} = 1$ and so $\beta = \frac{16}{9}$, leading to

$$F_s = \frac{32}{3}\pi\mu R_d v_d. \qquad (22)$$

From which we note the scaling:

$$F_s \sim R_d v_d. \qquad (23)$$

### $F_v$ – Drag Between the Moving Droplet and the Solid Oil-Water Interface

The motion of the droplet along the solid-like oil-water interface generates flows inside the droplet that oppose its motion. For the range of contact angles $\theta_c$ studied here (Supplementary Fig. S3), in which the interface is partially wet by the droplet, the magnitude of this drag has been shown to scale with the contact area between the droplet and the interface, i.e.[8–11]

$$F_v \sim R_d^2 v_d. \tag{24}$$

### Possible Scaling Relations

From Eqs. 16, 20, 23, and 24. we finally note three possible distinct scaling regimes:
1. An oil-water interface drag-dominated regime, in which $F_s \ll F_v$ and hence $v_d \sim R_d^4$ (corresponding to the scaling in our data, Fig. 2c).
2. A Strokes drag-dominated regime, in which $F_s \gg F_v$ and hence $v_d \sim R_d^5$.
3. An intermediate regime in which $F_s \approx F_v$ and the scaling exponent is between 4 and 5.

## Supplementary References

**Supplementary Videos**

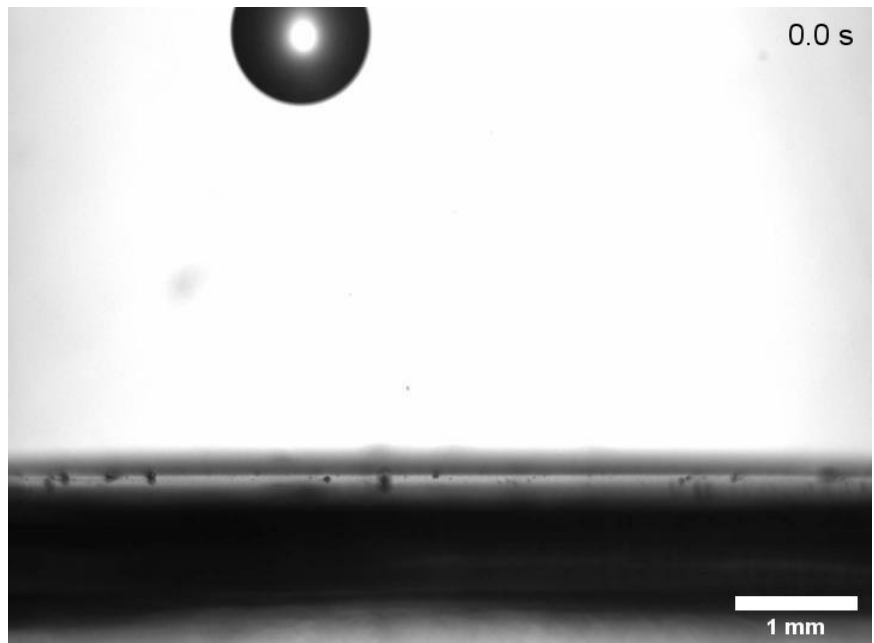

**Supplementary Video S1 | Placement of a water-ethanol droplet at a solid-like oil-water interface at which a layer of CNCS is assembled.** Droplet radius $R_\mathrm{d} = 0.8$ mm, droplet volume $V_\mathrm{d} = 2$ µL, ethanol volume fraction $\phi_\mathrm{EtOH} = 0.25$.

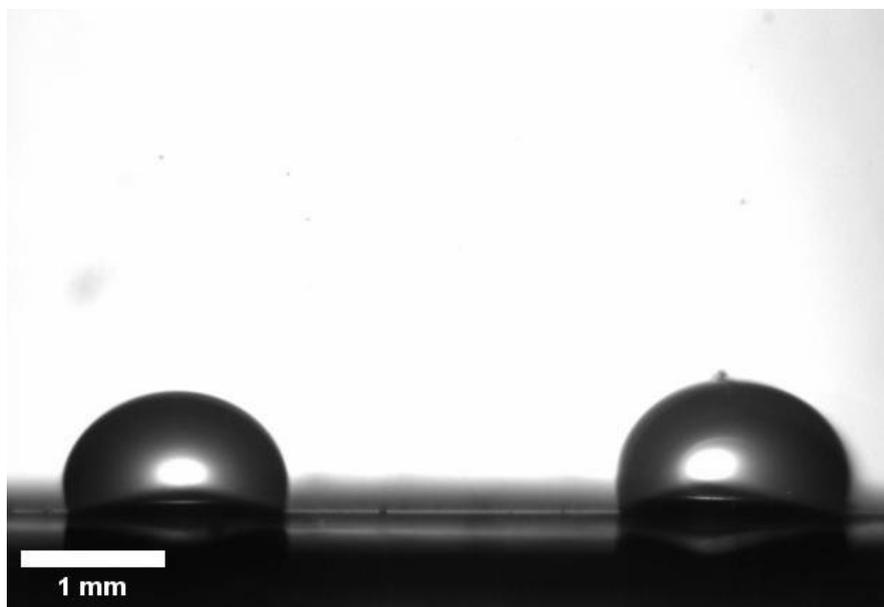

**Supplementary Video S2 | Placement of a second water-ethanol droplet at a solid-like oil-water interface.** Droplet radius, $R_\mathrm{d} = 0.8$ mm, droplet volume, $V_\mathrm{d} = 2$ µL, ethanol volume fraction, $\phi_\mathrm{EtOH} = 0.25$.

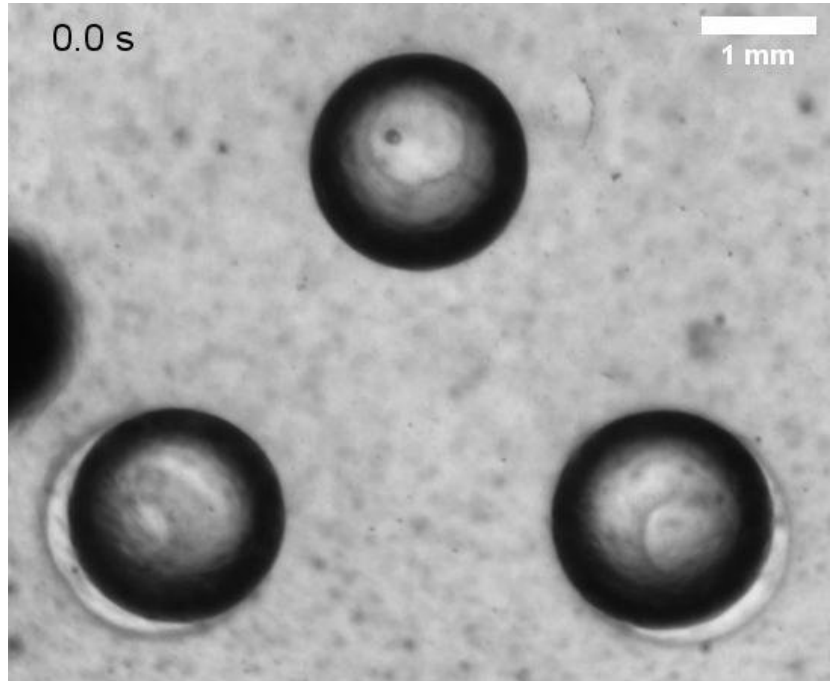

**Supplementary Video S3 | Attraction between three water-ethanol droplets, printed at a solid-like oil-water interface.** Droplet radius $R_d = 0.95$ mm, droplet volume $V_d = 2$ μL, ethanol volume fraction $\phi_{EtOH} = 0.5$. As in Fig. 1c, main text.

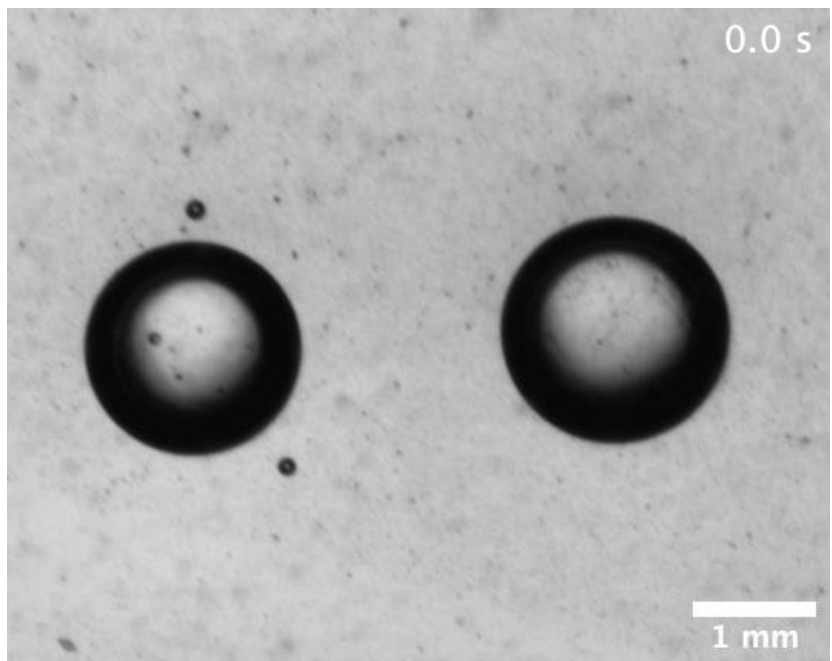

**Supplementary Video S4 | Attraction between two water-ethanol droplets manually placed at a solid-like oil-water interface.** Droplet radius $R_d = 0.95$ mm, droplet volume $V_d = 2$ μL ethanol volume fraction $\phi_{EtOH} = 0.5$.

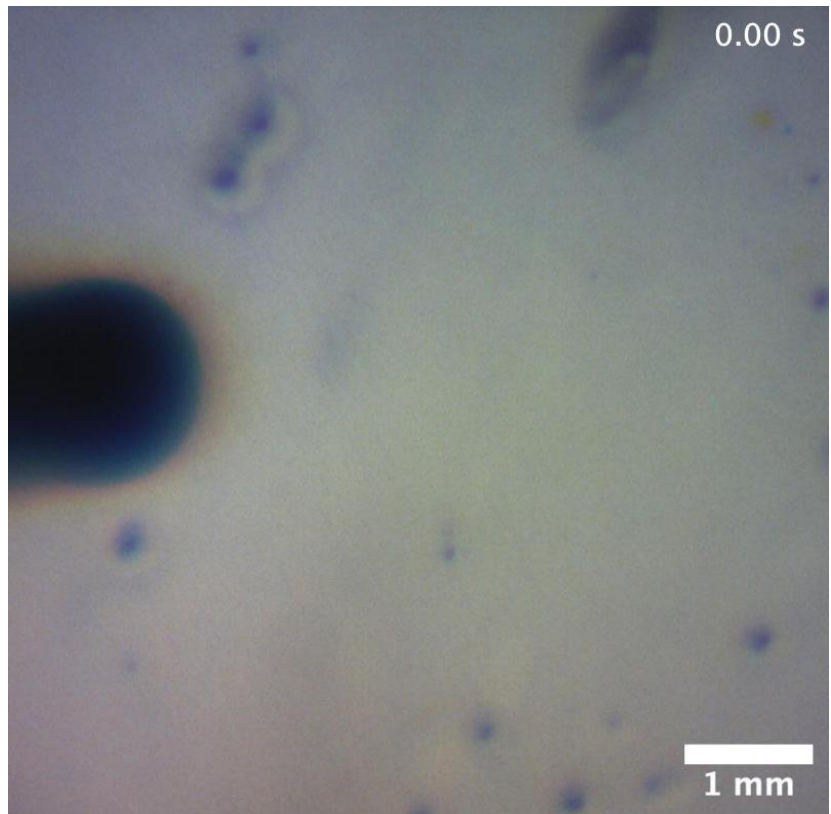

**Supplementary Video S5 | Printing and self-assembly of a two-dimensional hexagonal lattice of aqueous droplets at a solid-like oil-water interface.** Droplet volume $V_d = 0.5$ μL, ethanol volume fraction $\phi_{EtOH} = 0.25$. As in Fig. 3b.

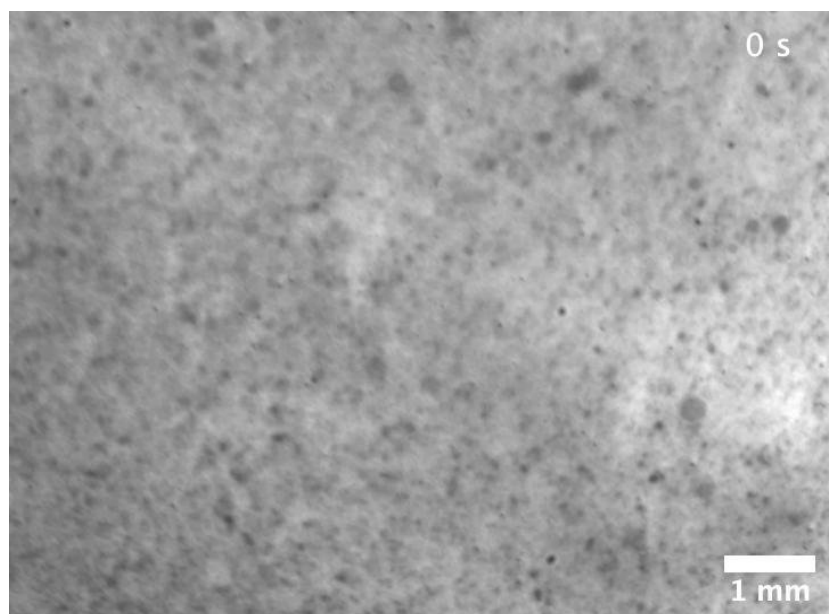

**Supplementary Video S6 | Manual placement and square lattice formation of bidisperse aqueous droplets at a solid-like oil-water interface.** Droplet volumes $V_d = 0.2$ μL and $V_d = 2$ μL, ethanol volume fraction $\phi_{EtOH} = 0.5$.

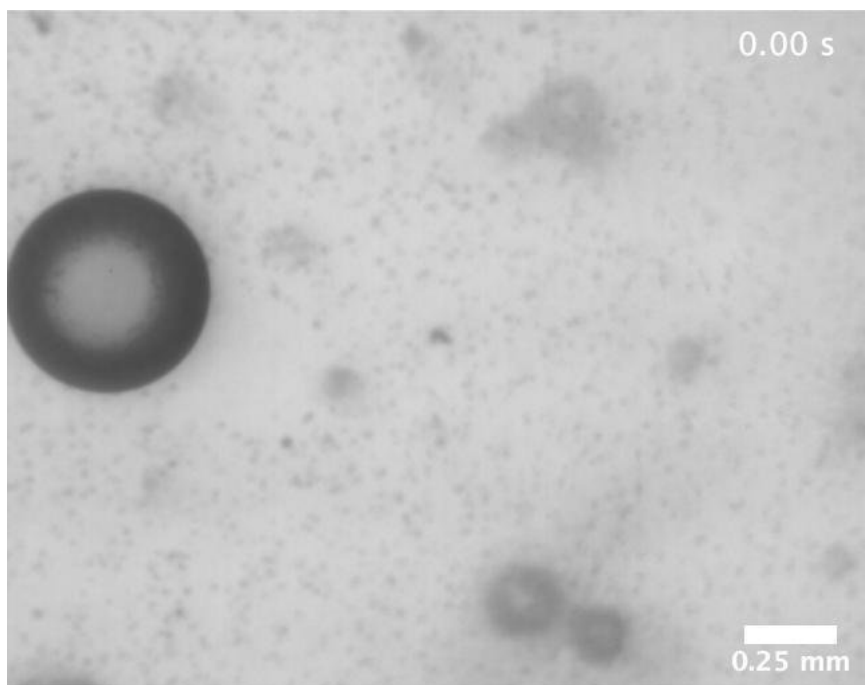

**Supplementary Video S7 | Thermoplasmonic manipulation of a droplet at a solid-like CNCS/AuNPS-covered oil-water interface.** A 532-nm continuous-wave laser is focussed at the solid-like oil-water interface leading to motion of the droplet. Droplet volume $V_d = 0.1$ μL, ethanol volume fraction $\phi_{EtOH} = 0.4$, laser power density $P \approx 160$ mW mm$^{-2}$. As in Fig. 5b.

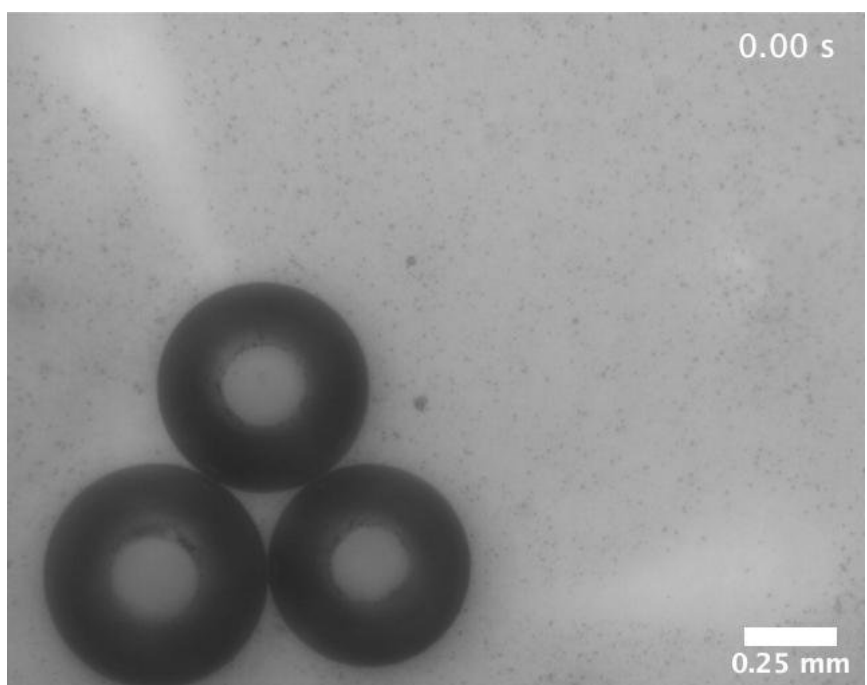

**Supplementary Video S8 | Thermoplasmonic manipulation of two droplets at a solid-like CNCS/AuNPS-covered oil-water interface.** A 532-nm continuous-wave laser, wavelength 532 nm, is focussed at the solid-like oil-water interface leading to the attraction of two droplets towards the laser and their subsequent assembly. Droplet volume $V_d = 0.1$ μL, ethanol volume fraction $\phi_{EtOH} = 0.4$, laser power density $P \approx 160$ mW mm$^{-2}$. As in Fig. 5d.